\documentclass[aps,twocolumn,superscriptaddress,showkeys,showpacs]{revtex4}
\usepackage{amssymb}
\usepackage{amsfonts}
\usepackage{amsmath}
\usepackage{amsthm}
\usepackage{epsfig}
\usepackage{color} 
\usepackage{subfigure}
\usepackage{bbm}

\begin{document}

\title{Chimera Death: Symmetry Breaking in Dynamical Networks}

\author{Anna Zakharova}
\affiliation{Institut f{\"u}r Theoretische Physik, Technische Universit\"at Berlin, Hardenbergstra\ss{}e 36, 10623 Berlin, Germany}
\author{Marie Kapeller} 
\affiliation{Institut f{\"u}r Theoretische Physik, Technische Universit\"at Berlin, Hardenbergstra\ss{}e 36, 10623 Berlin, Germany}
\author{Eckehard Sch{\"o}ll}
\email[corresponding author: ]{schoell@physik.tu-berlin.de}
\affiliation{Institut f{\"u}r Theoretische Physik, Technische Universit\"at Berlin, Hardenbergstra\ss{}e 36, 10623 Berlin, Germany}

\date{\today}

\begin{abstract}

For a network of generic oscillators with nonlocal topology and symmetry-breaking coupling we establish novel partially coherent 
inhomogeneous spatial patterns, which combine the features of chimera states (coexisting incongruous coherent and incoherent domains) and oscillation death (oscillation suppression), which we call {\it chimera death}. We show that due to the interplay of nonlocality and breaking of rotational symmetry by the coupling two distinct scenarios from oscillatory behavior to a stationary state regime are possible: a transition from an amplitude chimera to chimera death via in-phase synchronized oscillations, and a direct abrupt transition for larger coupling strength.

\end{abstract}


\maketitle

Spontaneous symmetry breaking in a complex dynamical system is a fundamental and universal phenomenon which occurs in diverse fields such as physics, chemistry, and biology \cite{TUR52}.
It implies that processes occurring in nature favor a less symmetric configuration, although the underlying principles can be symmetric. This intriguing concept has recently gained renewed interest generated by the enormous burst of works on chimera states and on oscillation death, which have emerged 
independently. In this Letter we draw a relation between these two.

Chimera states correspond to the situation when an ensemble of identical elements self-organizes into two coexisting and spatially separated domains with dramatically different behavior, i.e., spatially coherent and incoherent oscillations \cite{KUR02a,ABR04}. 
They have been the subject of intensive theoretical investigations, e.g. \cite{SET08,LAI09,MOT10,MAR10,OLM10,BOR10,SHE10,WOL11,OME11,LAI11,OME13,HIZ13}.
Experimental evidence of chimeras has only recently been provided for optical \cite{HAG12}, chemical \cite{TIN12}, mechanical \cite{MAR13} and electronic \cite{LAR13} systems. These peculiar hybrid states may also account for the observation of partial synchrony in neural activity \cite{OLB11}, like unihemispheric sleep, i.e., the ability of some birds or dolphins to sleep with one half of their brain while the other half remains aware \cite{RAT00,SMA12}. Chimera states have been initially found for phase oscillators \cite{KUR02a}, and they typically occur in networks with nonlocal coupling. Recently, for globally coupled oscillators it has been shown that such spatio-temporal patterns can be also connected to the amplitude dynamics (amplitude-mediated chimeras) \cite{SET13,SCH14a}. However, the global coupling topology does not provide a clear notion of space, which is crucial for chimera states. A further open question is whether chimeras can be extended to more general symmetry breaking states.

Another fascinating effect which requires the break-up of the system's symmetry as a crucial ingredient is oscillation death 
which refers to stable inhomogeneous steady states (IHSS) which are created through the coupling of self-sustained oscillators. This regime occurs when a homogeneous steady state splits into at least two distinct branches - upper and lower - which represent a newly created IHSS \cite{BAR85,TSA06,ULL07,KUZ04}. For a network of coupled elements oscillation death implies that its nodes occupy different branches of the IHSS. Oscillation death is inherent in various systems and its existence has been confirmed experimentally, e.g., in chemical reactors \cite{DOL88}, chemical oscillators \cite{CRO89}, chemical droplets \cite{TOI08}, electronic circuits \cite{HEI10} and thermokinetic oscillators \cite{ZEY01}. Moreover, it is especially significant in the light of applications to biological systems like neuronal networks \cite{CUR10}, genetic oscillators \cite{KOS10a}, calcium oscillators \cite{TSA06}, and it has been proposed as a basic mechanism for morphogenesis and cellular differentiation, for instance, stem 
cell differentiation \cite{SUZ11}.
 Oscillation death has been shown to exist for special time-delayed \cite{ZAK13a} and repulsive \cite{HEN13} types of coupling as well as coupling through conjugate \cite{ZOU13,KAR14} or dissimilar variables \cite{LIU12c}. Furthermore, in the context of network topologies, it has been studied for the two limit cases of global (all-to-all) and local 
coupling \cite{KOS10}. However, the influence of nonlocal coupling, which interpolates between these limit cases, on IHSS is still unresolved.

Therefore, chimera states and oscillation death have been previously investigated independently, although they possess a relevant common feature, both implying the break-up of symmetry in a dynamical network. Consequently, a significant and challenging task here is to provide systematically the bridging between these two phenomena, which will introduce more insight into the concept of symmetry-breaking in general, and fill the existing gaps in understanding chimeras and oscillation death in particular.

In this Letter, we investigate the relation between nonlocality of the network with symmetry-breaking oscillation death, and thus the formation of inhomogeneous steady states under these specific conditions. We establish a novel connection between chimera
states and oscillation death, which we call {\it chimera death}. This pattern generalizes chimeras to steady states, and it occurs in networks with nonlocal topology and a coupling which breaks the rotational symmetry.
The population of oscillators splits into distinct coexisting domains of (i) spatially coherent oscillation death (where neighboring oscillators populate essentially the same branch of the inhomogeneous steady state) and (ii) spatially incoherent oscillation death (where the sequence of populated branches of neighboring nodes is completely random in the inhomogeneous steady state). Moreover, in the presence of nonlocal coupling which allows to introduce a spatial scale in the system, providing the evident separation of spatial domains with different dynamical regimes, we find chimera behavior with respect to the amplitude, rather than the phase, i.e, {\it amplitude chimeras}. While increasing the coupling range for a fixed coupling strength, we uncover a novel scenario from amplitude chimera (where one sub-population is oscillating with spatially coherent amplitude, while the other displays 
oscillations with spatially incoherent amplitudes) to chimera death via an in-phase synchronized state. Additionally, we show that for larger values of coupling strength an abrupt transition from amplitude chimera to chimera death is possible with increasing coupling range.

We analyze the paradigmatic model of Stuart-Landau (SL) oscillators \cite{KUR02,ATA03,FIE09,CHO09,KYR13,SCH13b,POS13a,SCH14a},
\begin{equation}
  \label{eq:model}
  \dot{z}=f(z) \equiv (\lambda+i\omega - \left|z\right|^2)z,
\end{equation}
where $z = r e^{i \phi}=x+iy \in \mathbbm{C}$, $\lambda, \omega \in \mathbbm{R}$. For $\lambda>0$, the uncoupled system exhibits self-sustained limit cycle oscillations with radius $r_0=\sqrt{\lambda}$ and frequency $\omega$. Therefore, the Stuart-Landau system represents a generic model for nonlinear oscillators close to a Hopf bifurcation. We investigate a ring of $N$ nonlocally 
coupled Stuart-Landau oscillators:
\begin{equation}
   \label{eq:network}
   \dot{z}_j = f(z_j) + \frac {\sigma}{2P} \sum_{k=j-P}^{j+P} ( \mathrm{Re} z_k-  \mathrm{Re} z_j),
\end{equation}
where $j=1,2,...,N$.
The coupling parameters, which are identical for all links, are the
coupling strength $\sigma \in \mathbbm{R}$ and the coupling range $P/N$, where $P$ corresponds to the number of nearest neighbors in each direction on a ring. Here we consider coupling only in the real parts, since this breaks the rotational $S^1$ symmetry of the system which is a necessary condition for the existence of nontrivial steady states $z_j \neq 0$ and thus for oscillation death.
With respect to applications this means that the oscillators are coupled only through a single real variable $x$.

\begin{figure}[]
\begin{center}
\includegraphics[width=0.40\textwidth]{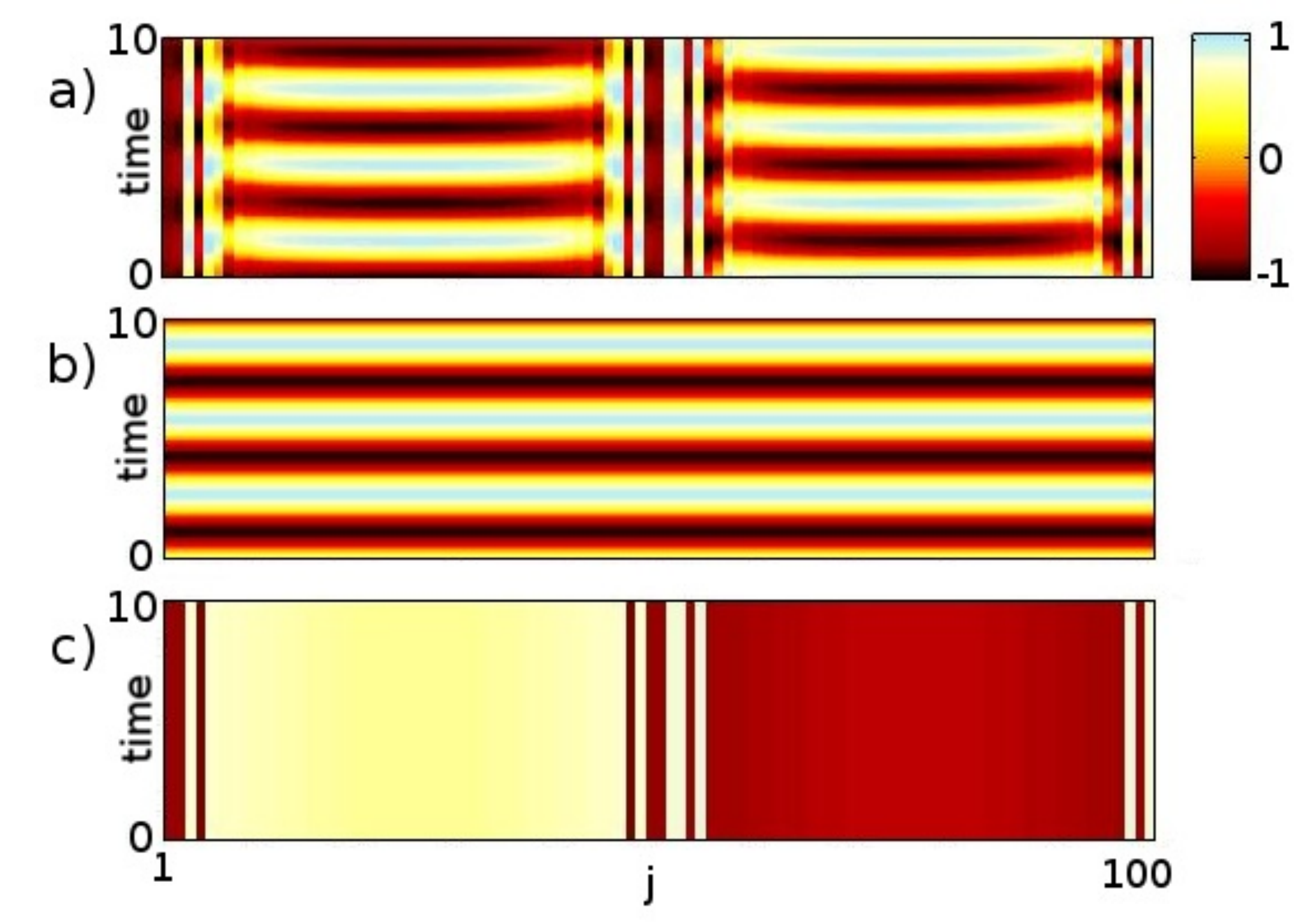}
\end{center}
\caption{Space-time plots for the variable $y_j(t)$ for coupling strength $\sigma=14$ and varying nearest neighbors number P. (a) $P=4$: amplitude chimera, (b) $P=5$: in-phase synchronization, (c) $P=33$: chimera death (coexistence of spatially coherent and incoherent oscillation death). Other parameters: $N=100$, $\lambda=1$, $\omega=2$. Transients of $t=1000$ were cut off. Initial conditions were chosen similar to \cite{OME11}.}
\label{fig:1}
\end{figure}

The interplay of two features which we introduce to the coupling, nonlocality and symmetry-breaking, gives rise to the three distinct regimes in Eq. (\ref{eq:network}) which are shown as space-time plots color-coded by the variable $y_j$ 
in Fig. \ref{fig:1}. In contrast to {\it amplitude-mediated chimeras} which have  recently been found for global coupling \cite{SET13}, we show here a pure amplitude chimera (Fig. \ref{fig:1}a) in the presence of nonlocal coupling, which allows for the definition of spatial distance in the system. Therefore, this regime is characterized by the coexistence of two distinct domains separated in space: one sub-population is oscillating with spatially coherent amplitude and the other exhibits oscillations with spatially incoherent amplitudes, i.e. the sequence of amplitudes of neighboring oscillators is spatially chaotic similar to chimeras found in coupled maps\cite{OME11}. The phases $\phi_j$ of neighboring oscillators, on the other hand, are in synchrony, in contrast to classical chimeras \cite{KUR02a,ABR04}. While varying the coupling range $P/N$ for fixed value of coupling strength $\sigma$ we observe a transition from amplitude chimeras to an in-phase synchronized state (Fig. \ref{fig:1}b). Further 
increasing the coupling range we find a novel regime, which combines the properties of chimera 
state and oscillation death. Therefore, we call it chimera death (Fig. \ref{fig:1}c). The system of identical oscillators breaks up into two sub-populations: (i) spatially coherent oscillation death, where the neighboring elements of the network populate the same branch of the inhomogeneous steady state, either $y^{*1} \approx +1$ or $y^{*2} \approx -1$, and (ii) spatially incoherent inhomogeneous steady states, where the sequence of states $y^{*1}, y^{*2}$ on the two branches for neighboring elements is completely random.

\begin{figure}[]
\begin{center}
\includegraphics[width=0.49\textwidth]{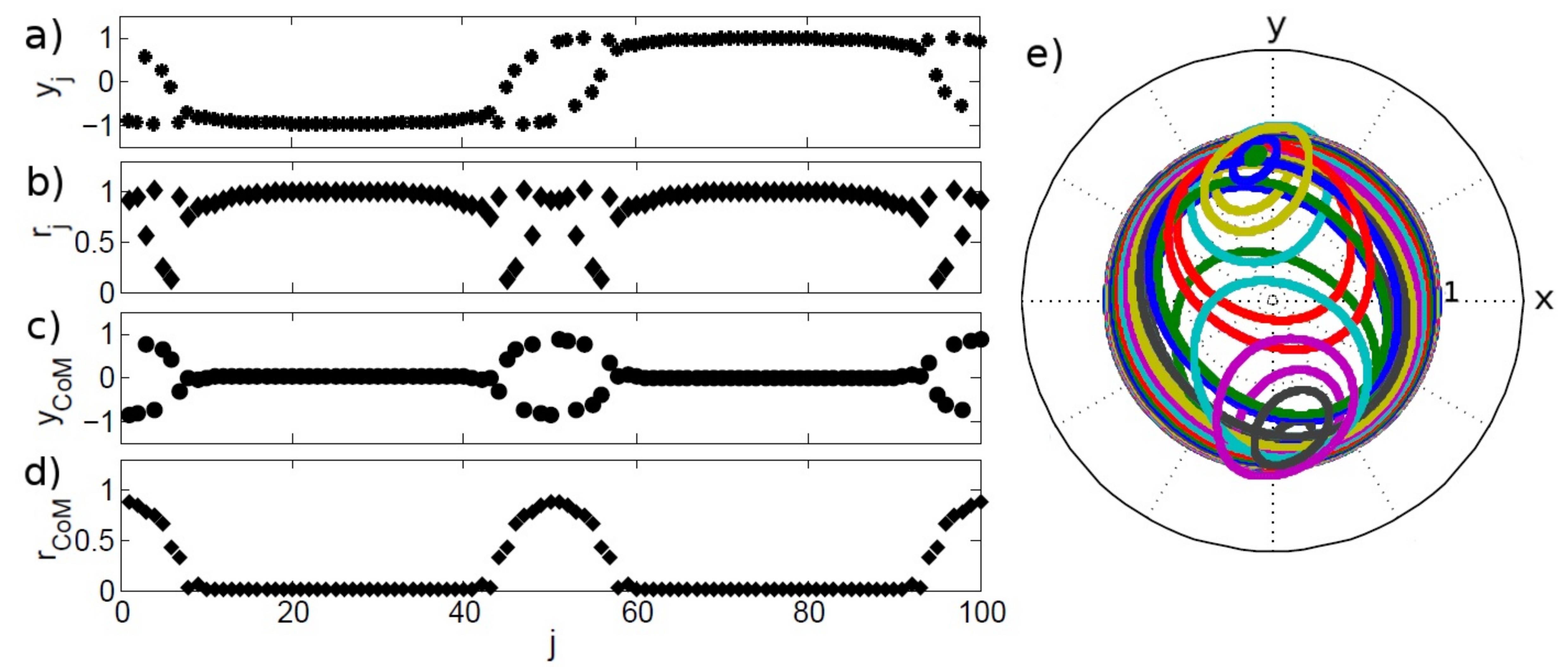} 
\end{center}
\caption{Amplitude chimera: Snapshots at $t=1000$ of (a) the variable $y_j$  and (b) the amplitude $r_j$ (with respect to the origin). Center of mass (CoM) averaged over one period of each oscillator for (c) the variable $y_{CoM}$, (d) the amplitude $r_{CoM}$ (i.e., the shift of the CoM from the origin). (e) Phase portraits of all oscillators in the complex $z=x+iy$ plane. In (c)-(e) transients of $t=1000$ were cut off. Other parameters: $N=100$, $P=4$, $\sigma=14$, $\lambda=1$, $\omega=2$.}
\label{fig:2}
\end{figure}

To provide more insight into the amplitude chimera state, we analyze snapshots of the variable $y_j$ (Fig. \ref{fig:2}a) and the amplitude $r_j$ at time $t=1000$. The coherent and incoherent domains are well pronounced, a typical feature of chimera states. The elements of the coherent part perform oscillations with nearly the same amplitude $r$, if measured from the origin (Fig. (\ref{fig:2})b). A relevant observation which is crucial for the understanding of the dynamics within the incoherent domain is that the center of mass of each oscillator is shifted away from the origin. To illustrate this, Fig. \ref{fig:2}c displays the $y$ coordinate of the center of mass $y_{CoM}=\int_0^T y_i(t) dt/T$, where $T=2\pi/\omega$ is the oscillation period, for each oscillator $j$. The elements of the coherent sub-population are characterized by zero shift of the center of mass $y_{CoM}=0$, therefore, they perform oscillations around the origin. In contrast to that, the incoherent part demonstrates a clear displacement of 
the centers of mass, and the central element exhibits the maximum drift from the 
origin. It is important to note that the mean phase velocity profile cannot serve as an indicator for the amplitude chimera, since the mean phase velocity is constant for all oscillators, in contrast to classical phase chimeras \cite{KUR02a}.
However, the distance between the center of mass of each oscillator and the origin $r_{CoM}$ plotted versus space $j$ discloses the characteristic signature of a chimera state (Fig. \ref{fig:2}d). Its shape strikingly resembles the typical phase velocity profile which is usually observed for coupled phase oscillators. Fig. \ref{fig:2}e shows the phase portraits of all oscillators in the complex $z$ plane clearly demonstrating the limit cycles with different amplitudes and centers of mass. An important observation is that amplitude chimeras also appear for local coupling.
This is not surprising, since for synthetic genetic oscillators the local coupling may generate partial oscillation death where some of the nodes are in the inhomogeneous steady state while others oscillate with different amplitudes \cite{KOS10,KOS13}.

\begin{figure}[]
\begin{center}
\includegraphics[width=0.49\textwidth]{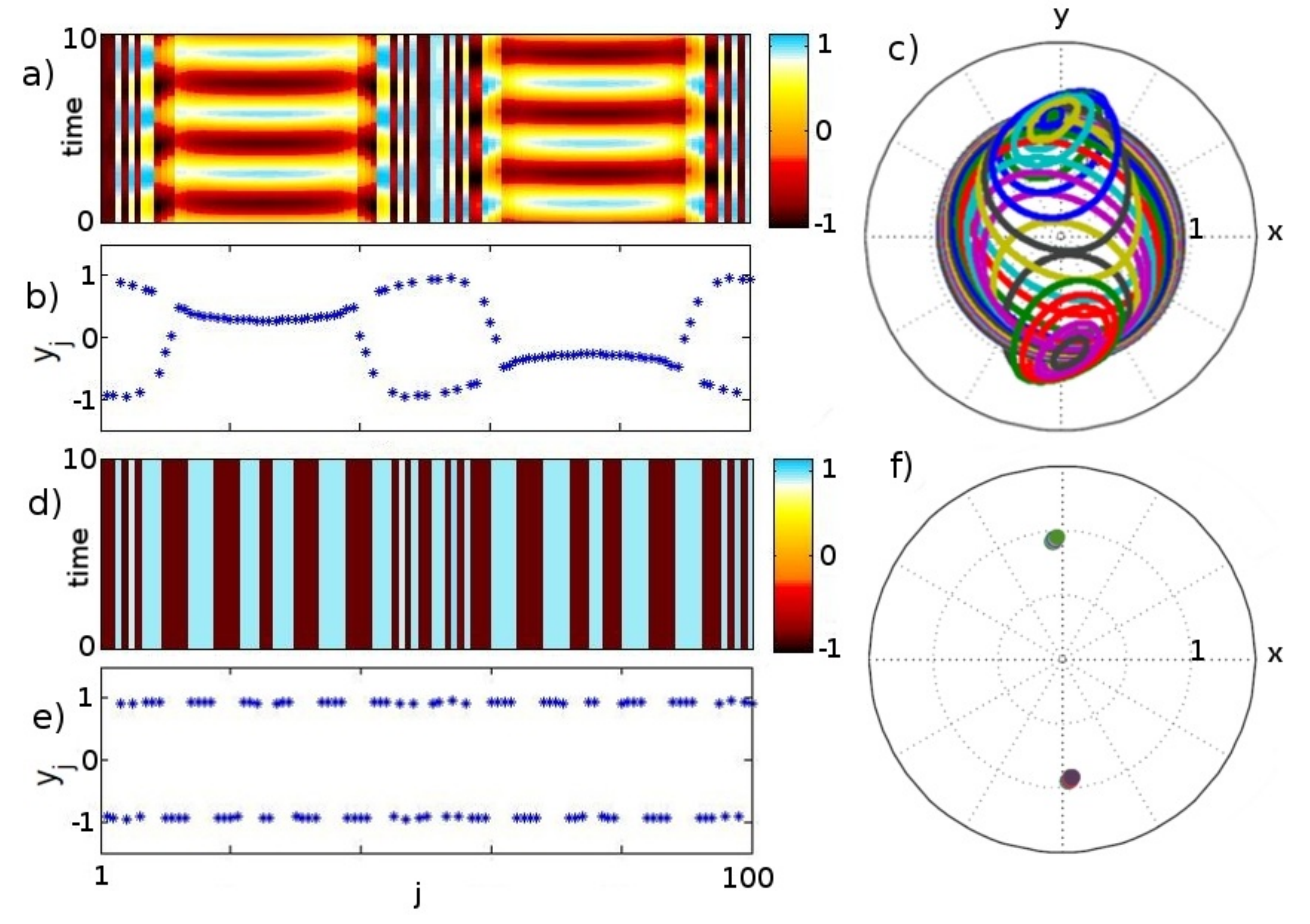}
\end{center}
\caption{Direct transition from amplitude chimera to chimera death for $\sigma=26$ by changing the coupling range. (a)-(c): Oscillating state (amplitude chimera) for $P=4$. (d)-(f): Stable inhomogeneous steady state (chimera death) for $P=5$. (a),(d): space-time plot for the variable $y_j$, (b),(e) snapshot for the variable $y_j$, (c),(f) phase portraits of all oscillators in the complex $z=x+iy$ plane. 
Transients of $t=1000$ were cut off. Other parameters: $N=100$, $\lambda=1$, $\omega=2$.}
\label{fig:3}
\end{figure}

Interestingly, for fixed value of the coupling range the incoherent domain of the amplitude chimera can be enlarged by raising the coupling strength. It can be clearly seen by pairwise comparison of space-time plots and snapshots for the same value of the coupling range $P/N=0.04$ and two different values of the coupling strength: Fig. \ref{fig:1}a, Fig. \ref{fig:2}a for $\sigma=14$ and Fig. \ref{fig:3}a, Fig. \ref{fig:3}b for $\sigma=26$, respectively. This observation is additionally illustrated by phase portraits: the number of elements oscillating with spatially incoherent amplitudes and with the center of mass shifted from the origin is smaller for the coupling value $\sigma=14$ (Fig. \ref{fig:2}e) if compared to the case of $\sigma=26$ (Fig. \ref{fig:3}c).

Surprisingly, the chimera death pattern also undergoes structural changes while tuning the coupling parameters, and we observe qualitatively new types of chimera death. In particular, if the value of coupling strength is fixed the variation of the coupling range leads to the formation of clusters within the coherent domains. Therefore, a chimera death state is not exclusively characterized by two uniform coherent domains as depicted in Fig. \ref{fig:1}c. On the contrary, if the number of nearest neighbors for every element in the network is decreased, the coherent domains split into multiple clusters. The maximum number of clusters observed is shown in Fig. \ref{fig:3}d-f including a typical phase portrait for the inhomogeneous steady state (Fig. \ref{fig:3}f).

\begin{figure}[]
\begin{center}
\includegraphics[width=0.40\textwidth]{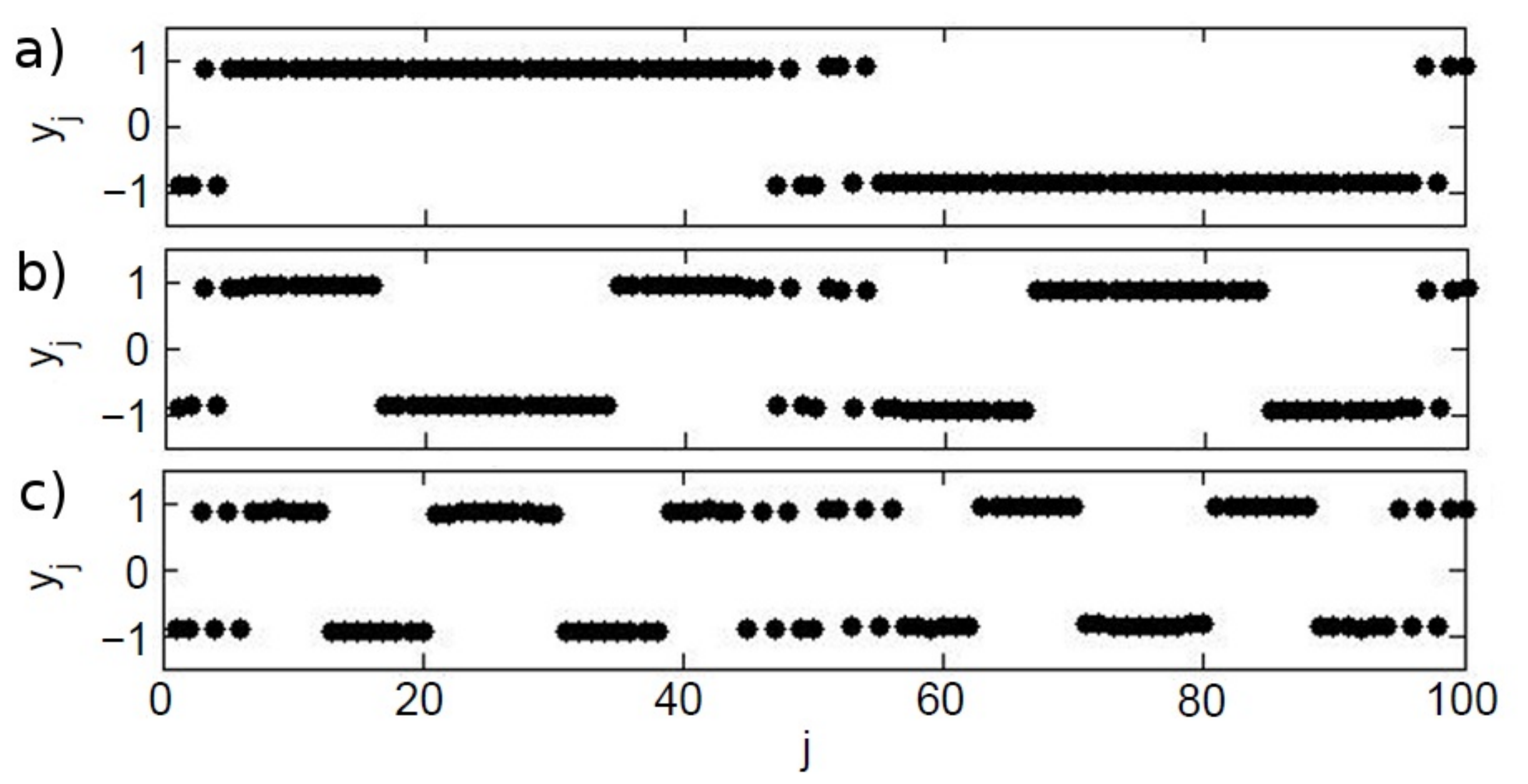}
\end{center}
\caption{Chimera death for fixed value of coupling strength $\sigma = 20$ and different values of coupling range $P/N$. Snapshots for the variable $y$: (a) $P = 40$, (b) $P = 20$, (c) $P = 11$.
Transients of $t=1000$ were cut off. Other parameters: $N=100$, $\lambda=1$, $\omega=2$.}
\label{fig:4}
\end{figure}

For the better understanding of the chimera death evolution we examine the snapshots of the variable $y$ for different values of the coupling range (Fig. \ref{fig:4}). 
The coherent domains of chimera death consist of one cluster in the state $y \approx +1$ and one cluster in the state 
$y \approx -1$ if the coupling range is large enough (e.g. $P=40$): the elements belonging to one domain populate the upper branch of the inhomogeneous steady state while the nodes from the other go to the lower branch (Fig. \ref{fig:4}a). Decreasing the coupling range ($P=20$) initiates the formation of clusters within either of the coherent domains, while the incoherent part remains at first unchanged. In more detail, central elements of the coherent sub-populations switch to the opposite branch of the inhomogeneous steady state: from the upper branch to the lower or vise versa (Fig. \ref{fig:4}b). This process is repeated while the coupling range is further decreased. Moreover, the incoherent domains are enlarged for the smaller coupling range ($P=11$) (Fig. \ref{fig:4}c). 

\begin{figure}[]
\begin{center}
\includegraphics[width=0.40\textwidth]{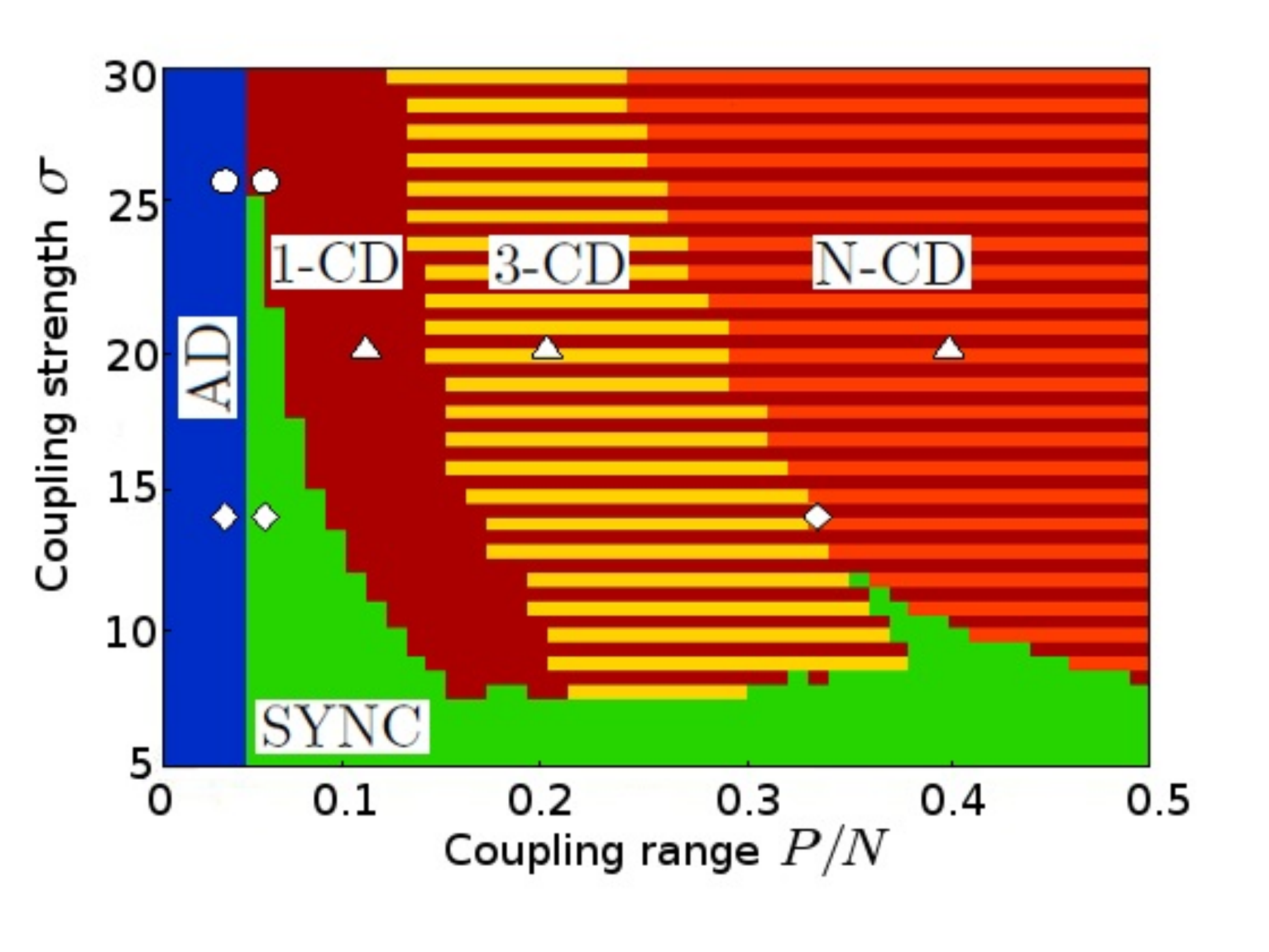}
\end{center}
\caption{Map of dynamic regimes for $N=100$ in the plane of coupling range $P/N$ and coupling strength $\sigma$. 1-CD: 1-cluster chimera death; 3-CD: 3-cluster chimera death; N-CD: multi-cluster ($>3$) chimera death. AC: amplitude chimera; SYNC: in-phase synchronized oscillations. 
Other parameters: $\lambda=1$, $\omega=2$. 
Diamonds, circles, and triangles mark the parameter values chosen in Figs. \ref{fig:1}, \ref{fig:3}, and \ref{fig:4}, respectively.}
\label{fig:5}
\end{figure}

To gain an overall view of the dynamics in the network of nonlocally coupled elements with symmetry-breaking coupling we fix the values of parameters $N=100$, $\lambda=1$, $\omega=2$ and vary the range $P/N$ and the strength $\sigma$ of the coupling. The oscillatory behavior of the system Eq. (\ref{eq:network}) is represented by amplitude chimera (blue (gray) region in Fig. \ref{fig:5}), which is observed for small coupling range $P/N<0.05$, and by in-phase synchronized oscillations (light green region in Fig. \ref{fig:5}), which arise beyond a threshold value $P/N=0.05$ and exist for relatively small coupling strength. The steady state solutions occur for larger values of coupling parameters and are manifested by chimera death (red (dark gray) region in Fig. \ref{fig:5}). The existence of two distinct transition scenarios from the oscillatory to the steady state regime becomes evident from Fig. \ref{fig:5}. For a small value of coupling strength ($\sigma=14$) the chimera death state is born from amplitude 
chimera by passing through in-phase synchronized oscillations when the coupling range is increased (diamonds in Fig. \ref{fig:5}). In contrast, for a large value of the coupling strength ($\sigma=26$) a slight increase of the coupling range from $P/N=0.04$ to $P/N=0.05$ destroys amplitude chimeras and directly leads to chimera death (circles in Fig. \ref{fig:5}). Moreover, the phase diagram shown in Fig. \ref{fig:5} clearly indicates the predominant role of oscillation death in the network's dynamics. The prevailing region of chimera death is divided into several regimes depending on the number of coherent clusters. Chimera death may consist of two coherent domains, which are formed by one cluster on the upper and one on the lower branch of the steady state (dark and light red stripes in Fig \ref{fig:5}). Moreover, the coherent regions may split into three clusters each (dark red and yellow stripes in Fig \ref{fig:5}) or more (red (dark gray) in Fig \ref{fig:5}). The typical snapshots in Fig \ref{fig:4}a,b,c 
provide additional illustration for different multiple chimera death states, and the corresponding coupling parameters are indicated in the phase diagram (Fig \ref{fig:5}) by triangles. 

In conclusion, we have related two different phenomena - chimera states and oscillation death - which have a common signature of symmetry breaking in dynamical networks, thus revealing two novel types of chimeras, i.e., coexisting incongruous spatially coherent and incoherent states: amplitude chimera and chimera death. These two findings generalize chimeras to amplitude dynamics and to inhomogeneous steady states, respectively. 
Moreover, we have uncovered the transition scenarios between these partially coherent spatio-temporal patterns. In particular, we have demonstrated that a network of identical elements with nonlocal symmetry-breaking coupling can be driven out of the oscillatory regime of amplitude chimera by two distinct mechanisms: either via passing through an in-phase synchronized state, a scenario which occurs at low coupling strength, or at sufficiently large coupling strength by an abrupt suppression of oscillations and therefore, the transition to chimera death, by increasing the coupling range. Moreover, transitions from in-phase synchronized oscillations to inhomogeneous stationary patterns with various numbers of coherent and incoherent clusters (chimera death) can be induced by increasing the coupling strength.
These findings may deepen our general understanding of partial synchronization patterns in complex networks.
Additionally, we have made a first important step towards the systematic study of oscillation death in the context of complex network topologies with nonlocal interactions of variable range and symmetry-breaking coupling.
By bridging between chimera states and oscillation death we hope to initiate numerous further questions, which may open up a new field of investigation, e.g., ``multi-cluster'' chimera death states, scenarios between various ``n-cluster'' states, or multiplicity of oscillation death branches for complex coupling topologies.

Since we have studied a generic model of coupled self-sustained oscillators, our results can be applied to a wide class of systems ranging from laser models, communication networks and power grids to biological networks. We believe that they are of particular importance for synthetic biology. For instance, while building synthetic circuits one can initiate the transitions between different regimes of operation, e.g., from self-sustained oscillations (amplitude chimera) to oscillation suppression (chimera death) by adjusting the architecture of the network (tuning coupling range and coupling strength) without changing the local dynamics (tuning parameters of the individual nodes). Moreover, our findings may provide a recipe for engineers when constructing a network where oscillation suppression in a specific configuration (chimera death with a certain number of cluster in the coherent domain) is desired.

We acknowledge Aneta Koseska, Isabelle Schneider, Katharina Krischer, and Bernold Fiedler for valuable discussions.   
This work was supported by DFG in the framework of SFB 910.

\bibliographystyle{prwithtitle}


\end{document}